\documentclass[twocolumn,aps,epsfig,graphics,]{revtex4}
\usepackage[dvips]{graphics}
\newcommand{\be}{\begin{equation}}
\newcommand{\ee}{\end{equation}}

\newcommand{\bea}{\begin{eqnarray}}
\newcommand{\eea}{\end{eqnarray}}
\newcommand{\bd}{\begin{displaymath}}
\newcommand{\ed}{\end{displaymath}}
\newcommand{\bi}{\begin{itemize}}
\newcommand{\ei}{\end{itemize}}
\newcommand{\bc}{\begin{center}}
\newcommand{\ec}{\end{center}}
\newcommand{\bfl}{\begin{flushleft}}
\newcommand{\efl}{\end{flushleft}}
\newcommand{\bfr}{\begin{flushright}}
\newcommand{\efr}{\end{flushright}}

\def\6{\partial}

\def\={\!\!\!&=&\!\!\!}
\def\+{\!\!\!&&\!\!\!+~}
\def\-{\!\!\!&&\!\!\!-~}

\begin{document}
\title{Temperature effect in the conductance of hydrogen molecule}
\author{M. Crisan and I. Grosu}
\affiliation{Department of Theoretical Physics, University of
Cluj, 3400 Cluj-Napoca, Romania}

\begin{abstract}
We present a many-body calculation for the conductance  of a conducting bridge of a simple hydrogen molecule
between $Pt$ electrodes.The  experimental results showed that the conductance $G=dI/dV$
has the maximum value near the quantum unit $G_{0}=2e^{2}/h$.
 The $I-V$ dependence presents peak and dip and we consider that the electron-phonon interaction is responsible
  for this behavior.
 At $T=0$ there is a step in this dependence for  the energy of phonons $\omega_{0}$
 which satisfies $eV=\omega_{0}$. We calculated the conductance  at finite temperature
 and showed that $dG(T)/dV\propto 1/4T\cosh^{2}\frac{eV-\omega_{0}}{2T}$.

\end{abstract}

\maketitle

\section{Introduction}

 The recent advances in manipulation of single molecules  permit to measure the transport
 properties of a setup formed from an individual molecule between two electrodes. In contrast to
 the quantum dots,  based on the semiconducting islands, the molecular devices have a more complicated electronic structure.

  In a remarkable experiment\cite{sm} a setup consisting from a single hydrogen molecule between Pt electrodes was measured
  and the conductance behavior demonstrated  the influence of the vibrational degree of freedom
  in the transport. The effect was also observed in various organic molecules \cite{zh}, carbon nanotubes \cite{le}
  fulerenes \cite{pa} but a theoretical description of these systems is more complicated because of the energy  spectrum of
  the electrons from these molecules.
   The simplest model which is realistic for the hydrogen molecule setup is to consider the coupling between the vibrational
   mode (considered as phononic) of frequency $\omega_{0}$ and electrons from the leads. The main effect which
   appears in the I-V characteristic is the occurrence of a step at $V= \omega_{0}/e$, which corresponds
   to dip versus peak in $d^{2}I/dV^{2}$. Such a behavior has been studied also in \cite{mi,ga1,ga2}.  Many
   models have been proposed by different authors \cite{co, la, za}in order to explain the influence of vibrational modes
   on the transport, but only recently Egger and Gogolin \cite{go} presented an analytical calculation explaining the
   current-voltage relation. Their calculation, which gives the correction to the current $\delta I$ ,is a perturbative calculation
   for  the  electron-phonon interaction and takes only the $g^{2}$ contributions similar to the approximation from Ref.\cite{fr, en}
   where the electron-phonon interaction is treated in the weak coupling approximation.  In a higher order of perturbation theory $g^{n}$
   the step behavior is expected at $n\omega_{0}$ .
   The occurrence of  the  step feature  at multiples of $\omega_{0}$ is related to the strong- coupling picture
   using the polaron transformation \cite{al}.

  In this paper we present a many- body calculation of the finite temperature conductance for the  hydrogen molecule
  between metallic leads. The  procedure is similar to this from Ref. [12] but we calculate the conductance of the setup
  close to the characteristic frequency $\omega_{0}$ at finite temperature. The peak (dip)which appear at this frequency
  has a finite width, due to thermal effect.

    The paper is structured as follows. In Sec.II we present the model and calculate the Green functions.
    In Sect.III we calculate the lowest order correction $\delta I$ to the current which is given by the electron
    -phonon processes. The concluding remarks are presented in Sec.IV.

  \section{Model and Green function}
We start with the Hamiltonian which describes the interaction
between electrons and the Holstein phonos used in Ref. \cite{ga2}
for the molecular dots  and  in Ref.\cite{fr} for study of the
inelastic scattering influence on the  the tunnelling current in
the two dimensional systems.The simple Hamiltonian, which is analytically
tractable, has the form :
\begin{equation}
H=H_{0}+H_{i}
\end{equation}
where
\begin{equation}
H_{0}= \epsilon_{0}d^{\dag}d+\omega_{0}b^{\dag}b
+\sum_{k\alpha}(\epsilon_{k}-\mu_{\alpha})c^{\dag}_{k\alpha}c_{k\alpha}
\end{equation}
\begin{equation}
H_{i}=\sum_{k,\alpha}(V_{\alpha}d^{\dag}c_{k\alpha}+H.c)+gQd^{\dag}d
\end{equation}
where $Q=b+b^{\dag}$.

In the Hamiltonian (1) $d$ and $d^{\dag}$ are the operators for
the single level $\epsilon_{0}$, the Holstein phonons with energy
$\omega_{0}$ are described by the operators $b^{\dag}$ and $b$, the
electrons with the energy $\epsilon_{k\alpha}$,($\alpha=L,R$) and
chemical potential $\mu_{\alpha}$ are described by the operators
$c_{k\alpha}$ and $c^{\dag}_{k \alpha}$. The interaction between the
electrons from the leads and the impurity is $V_{\alpha}$ and $g$
describes the interaction between the  localized electronic level
$\epsilon_{0}$ and the phonons. This Hamiltonian has been used by
Egger and Gogolin \cite{go} in this problem at $T=0$.Using the
equation of motion method we calculate the Green function
$G^{r}_{0}= <<d|d^{\dag}>>$ as
$G^{r}_{0}(\omega)=(\omega-\epsilon_{0}+i\Gamma)^{-1}$ where we define
$\Gamma = \Gamma_{L}+\Gamma_{R}$ and
$\Gamma_{\alpha}=\pi N(0)|V_{\alpha}|^{2}$.
The Green function describing the system is :
\begin{equation}
G^{r}(\omega)=G^{r}_{0}(\omega)+G^{r}(\omega)\Sigma^{r}(\omega)G^{r}_{0}(\omega)
\end{equation}
where the self energy $\Sigma^{r}(\omega)$ is taken in the lowest order and has the form\cite{en}:
\begin{widetext}
\begin{equation}
\Sigma^{r}(\omega)=-\frac{g^{2}}{2}\int d\omega'd\omega''
\frac{[Im G_{0}(\omega',\epsilon_{0})ImD_{0}(\omega'',\omega_{0})]A(T,\omega', \omega'')}
{\omega'+\omega''-\omega}
\end{equation}
\end{widetext}
where $A(T,\omega', \omega'')=1-f_{F}(\omega')+n_{B}(\omega'')$, $f_{F}(\omega)$ is the Fermi
function and $n_{B}(\omega)$ is the Bose function , and the  phononic Green function $D_{0}(\omega,\omega_{0})$
is :
\begin{equation}
D^{R}_{0}(\omega,\omega_{0})=\frac{1}{2\omega_{0}}\left[\frac{1}{\omega-\omega_{0}+i\delta}+\frac{1}{\omega+\omega_{0}+i\delta}\right]
\end{equation}
From Eqs.(5-6)we obtain :
\begin{equation}
Im\Sigma^{r}(\omega)=- g^{2}\sum_{\alpha,s=\pm 1}\frac{\Gamma_{\alpha} f_{F}(\omega_{0}-s(\bar\mu_{\alpha}-\omega))}
{(\omega+s\omega_{0})+\Gamma^{2}}
\end{equation}
where $\bar\mu=(\mu_{R}+\mu_{L})/2-\epsilon_{0}$ and $\bar\mu_{\alpha=L/R=\pm 1}=\bar\mu \pm eV/2$.

This  contribution is dominant  in the  correction in current for small  $\omega_{0}/\Gamma $ \cite{go}, and we will calculate
only this contribution at finite temperature.

\section{Current and conductance}

The electrical current through the dot can be calculated from the Green function $G^{r}(\omega)$ as :
\begin{equation}
I(V)=-\frac{4e}{h}\frac{\Gamma_{L}\Gamma_{R}}{\Gamma}\int d\omega [f_{L}(\omega)-f_{R}(\omega)]Im G^{r}(\omega)
\end{equation}
and for the case of $g=0$ we get the current $I_{0}$ as
\begin{equation}
I_{0}(V)=\frac{e}{h}\frac{4\Gamma_{L}\Gamma_{R}}{\Gamma}[\arctan(\bar\mu_{L}/\Gamma)-\arctan(\bar\mu_{R}/\Gamma)].
\end{equation}
For $V\rightarrow 0$ the transparency of the junction, $\Upsilon= (h/e^{2})dI/dV $ is
\begin{equation}
\Upsilon=\frac{4\Gamma_{L}\Gamma_{R}}{\Gamma}\frac{1}{1+(\bar\mu/\Gamma)^{2}}\leq 1.
\end{equation}
Using Eqs.(4-8) we calculate the correction to the current given by the inelastic
electron-phonon scattering  \cite {go} as:
\begin{equation}
\delta I_{inel}=\frac{e}{h}\frac{\Gamma_{L}\Gamma_{R}}{\Gamma}\int_{\bar\mu_{R}}^{\bar\mu_{l}}d\omega F(\omega, V)
\end{equation}
where $F(\omega)$ is given by :
\begin{equation}
F(\omega,V)=-\frac{\Gamma^{2}-\omega^{2}}{(\omega+\Gamma^{2})^{2}}\sum_{\alpha, s}g^{2}\Gamma_{\alpha}
\frac{f_{B}[\omega_{0}-s(\bar\mu_{\alpha}-\omega)]}{(\omega+s\omega_{0})+\Gamma^{2}}.
\end{equation}
 In order to calculate the  conductance $G=dI/dV$ and its derivative $dG/dI$ we will use the
 relation:
\begin{equation}
\frac{d}{dV}\int_{\bar\mu_{R}}^{\bar\mu_{L}}F(\omega, V)=\frac{e}{2}\left[F(\bar\mu_{L},V)+F(\bar\mu_{L},V)\right]
+\int_{\bar\mu_{R}}^{\bar\mu_{L}}d\omega \frac{dF}{dV}.
\end{equation}
Using this formula we calculate the derivative $d\delta I_{inel}/dV$ as :
\begin{equation}
\frac{d\delta I_{inel}}{dV}=-\frac{e^{2}}{h}\frac{\Gamma_{L} \Gamma_{R}}{\Gamma}g^{2}\left[f_{F}(\omega_{0}-eV)S_{1}
+\Theta(V-\frac{\omega_{0}}{e})S_{2}\right]
\end{equation}
where :
\begin{equation}
S_{1}= \sum_{\alpha=\pm 1}\frac{\Gamma_{\alpha}(\Gamma^{2}-\bar\mu^{2}_{-\alpha})}{(\Gamma^{2}+\bar\mu^{2}_{-\alpha})[\Gamma^{2}+(\bar\mu_{-\alpha}+
\alpha \omega_{0})^{2}]}
\end{equation}
and
\begin{equation}
S_{2}=\sum_{\alpha=\pm 1}\frac{\Gamma^{2}-(\bar\mu_{\alpha}-\alpha\omega_{0})^{2}}{(\bar\mu^{2}_{\alpha}+
\Gamma^{2})[(\bar\mu^{2}_{\alpha}-\alpha\omega_{0})+\Gamma^{2}]^{2}}.
\end{equation}
Using these results we calculate the contribution of the inelastic scattering in $\frac{dG}{dV}$ as :
\begin{equation}
\frac{d^{2}\delta I_{inel}}{dV^{2}}=-\frac{e^{2}}{h}\frac{\Gamma_{L}\Gamma_{R}}{\Gamma}g^{2}\frac{S_{1}+S_{2}}{4T\cosh^{2}
\left(\frac{eV-\omega_{0}}{2T}\right)}.
\end{equation}

This is the main result of this paper, which shows that at finite temperature the peak (dip) in the conductance derivative
$dG/dV$ has a finite width at $eV=\omega_{0}$ at low , but finite temperature. At $T=0$ we reobtain the $\delta(eV-\omega_{0})$
behavior predicted in Ref.[12].

 The contribution of the elastic scattering present a logarithmic divergence at $eV=\omega_{0}$ which create
 symmetric dip or peak in the differential conductance. The relative importance of the inelastic versus quasi-elastic
 contributions has been analyzed in \cite{go} at  $T=0$ where was showed that at $\omega_{0}/\Gamma <<1$ and large $\bar\mu$
 the inelastic channel is dominant. This result remain valid also at finite temperature and in the following we will discuss the
 conditions for the occurrence of a dip or peak in the inelastic correction. First we mention that this correction is not any more
 singular, as at $T=0$, but we obtain the result from \cite{go} in this limit. For $\Gamma_{L}=\Gamma_{R}$ and $\bar\mu=0$
  we have $\Upsilon=1$ for $\omega_{0}>2\Gamma$ instead of a dip we get a peak.This is a particular case , and for $\bar\mu\neq 0$
 and $\bar\mu>\sqrt{\Gamma^{2}+\omega_{0}^{2}/4}$ we have a peak.We also have at $\bar\mu=\pm \Gamma$ the transparency $\Upsilon=1/2$
 which is in fact the point of the peak-dip transition.

At $T=0$ the only way to obtain a finite life-time of phonons , which may give a smearing of the step and/or peak feature
is to include the electronic polarization in the Green function of the phonons, as was suggested in \cite{go}. However, such a
calculation implies the higher-order perturbations in $g$ and  it is difficult to be performed analytically. In this paper the discussion
concerning the occurrence of the step or/and peak is identically to that of the authors of \cite{go}, but the coupling to
the thermal phonons generate  in a natural way a smearing of the step or/and peak structure.

\section{Concluding Remarks}

 We analyzed , using the many body method, the transport in the simplest molecular dot consisting from a hydrogen molecule
 between the Pt leads. The analytical calculations of the effect given by the electron-phonon interaction on current at
 finite temperature have been performed.The obtained results can be regarded as complementary to the $T=0$ similar calculations
 presented in Ref.\cite{go}. However, at zero temperature the concept of phonon is not defined and in the transport we
 use the concept of inelastic scattering.On the other hand the experimental results showed the existence of behavior
  in the $dG/dI$ which cannot be described by the simple $\delta(eV-\omega_{0})$ behavior, which is specific for
  $T=0$. We showed that at finite temperature a dip or a peak in this quantity is described by
  $dG/dI=C(\bar\mu, \Gamma,\omega_{0})1/(4T\\cosh^{2}\frac{eV-\omega_{0}}{2T})$.This behavior is given by the inelastic scattering
  between electrons and phonons, the elastic contribution giving a non relevant contribution to the transport
  in this system, which can be considered as a molecular quantum dot. Our results at $T=0$ are identically with the
  results from \cite {go},and the signature the constant $C({\bar\mu, \Gamma,\omega_{0}})$ gives the same conditions.
 Our calculations completed  the microscopic model   presented in Ref.\cite{go}, but we consider that
 the transport in more complicated molecules is difficult to be treated analytically.However, the model can be a starting
 point for the study of transport in the more complex molecular systems.

 We thank Alexander Gogolin for useful correspondence on the subject.

\end{document}